# Intercity Connectivity and Urban Innovation

**Xiaofan Liang**\*, University of Michigan, xfliang@umich.edu

**César A. Hidalgo**, Center of Collective Learning, ANITI, TSE, IAST, IRIT, University of Toulouse

**Pierre-Alexandre Balland**, Utrecht University

**Siqi Zheng**, Massachusetts Institute of Technology

**Jianghao Wang**, Chinese Academy of Sciences

## Abstract

Urban outputs, from economy to innovation, are known to grow as a power of a city's population. But, since large cities tend to be central in transportation and communication networks, the effects attributed to city size may be confounded with those of intercity connectivity. Here, we map intercity networks for the world's two largest economies (the United States and China) to explore whether a city's position in the networks of communication, human mobility, and scientific collaboration explains variance in a city's patenting activity that is unaccounted for by its population. We find evidence that models incorporating intercity connectivity outperform population-based models and exhibit stronger predictive power for patenting activity, particularly for technologies of more recent vintage (which we expect to be more complex or sophisticated). The effects of intercity connectivity are more robust in China, even after controlling for population, GDP, and education, but not in the United States once adjusted for GDP and education. This divergence suggests distinct urban network dynamics driving innovation in these regions. In China, models with social media and mobility networks explain more heterogeneity in the scaling of innovation, whereas in the United States, scientific collaboration plays a more significant role. These findings support the significance of a city's position within the intercity network in shaping its success in innovative activities.

**Introduction**

In recent years, scholars have used the scaling framework to uncover several important relationships between city size and social and economic outcomes. In per capita terms, larger cities are more innovative, produce more output, and generate more employment[1-3]. They are also home to greater concentrations of complex and innovative economic activities[4, 5]. Nevertheless, these agglomeration effects cannot be understood as a function of city size alone. Cities are not isolated entities but nodes in networks involving the exchange of goods, people, and ideas[6-8]. A more nuanced understanding of agglomeration effects therefore requires consideration of the network dimension.

Research in recent decades has advanced our understanding of intercity connectivity and its impact on urban innovation. Building on Castell's notion of "a space of flows"[7-9], scholars have mapped intercity networks using data on commercial flights[10], shipping[11], rail transport[12], human mobility[13], and firm mobility[7]. For example, cities such as London and New York have been identified as "Alpha" cities in the network of global firms[14, 15], and Beijing as the quintessential center of mobility in China[16]. These observations are consistent with theories and studies that conceptualize cities as economic actors situated in social and institutional relationships to transfer resources, complement functions, and intervene in opportunities[17-19]. Many historical cities, such as Moscow, thrive because of their central influence and accessibility (i.e., closeness and betweenness centrality)[20]. For modern cities, knowledge is the most valuable resource. Here we find theories that emphasize the role of geographic proximity and knowledge spillovers[21-23] and empirical studies that show the effect of social, spatial, and academic connections on firm growth and innovation. For example, Bailey et al.[24] found that social media connections between counties predicted the likelihood of patent citations between counties, even after controlling for technology classes and geographic distance. Wang et al.[25] found that high-speed rail development (i.e., mobility connectivity) positively affects the growth of knowledge-intensive industries in China. Hohberger et al.[26] reported that scientific collaboration with universities and allies helps biotech firms stay close to the innovation center of the field.

Studies integrating intercity connectivity into the scaling framework are just beginning to develop. Early discussions of networks in the scaling literature are limited to networks that are internal to cities, such as the transportation networks defined by a city's infrastructure or the social networks that emerge in cities with multiple-story buildings[2, 27, 28], but have not yet incorporated intercity networks. For example, Bettencourt[2] models cities as space-filling fractals, with a dimension less than two because of the gaps generated by voids and empty spaces. Ribeiro et al. focus on the decay of human interactions with distance[27]. Whereas Molinero and Thurner continue this tradition with a model that explains scaling coefficients from a the geometric properties of a city[28]. While this line of research has helped us improve our understanding of urban scaling phenomena, it focuses on within-city networks and could benefit from data about a city's position in intercity networks. More recently, scholars have started to examine the impact of intercity connectivity on the scaling of urban performance. For example, Keuschnigg et al. found that cities' local attractivity (i.e., migration of educated and talented people from smaller places), rather than their sizes, explains the much of Swedish cities' superlinear scaling in wages[29]. The elasticity between wages and city sizes is further reduced when the scaling relationship controls for population composition (e.g., educational attainment, cognitive ability, and creative job characteristics etc.). The increasing returns to scale in urban wealth are also only observed in cities whose population and incoming commuters exceeding a well-defined threshold[30]. Only cities in a dominant position within an urban network can expect robust trajectories of superlinear growth[31]. Lei et al. reported that, as compared with population, interurban mobility interactions in China have greater impacts on cities' exports, which depend on intercity networks, than GDP, wages, or consumptions, which couple more closely with population[32]. Bonaventura et al. found that node centrality in the U.S. workforce mobility network outperforms population in explaining city innovation performance[33]. These studies further influence that population and intercity connectivity may affect the scaling of different urban outcomes.

We found that existing attempts to integrate intercity connectivity in the scaling literature are limited by the scale (small), type (mobility networks only), and geography (one country) of the intercity networks involved and by a narrow focus on interpreting scaling exponents. Intercity network data are, by definition, quadratic in the number of cities. Thus, unlike data on simpler urban characteristics, such as population, GDP, or education levels, they require dyadic data that are often unavailable in city-level statistics. Moreover, since each city may belong to multiple networks, we can expect each of these networks to have a different impact on specific technologies and industries. For example, transportation networks can be key for logistic-intensive industries while social networks may be more important for activities that involve the flow of ideas rather than. Intercity connectivity may also play a different role in developed versus developing urban contexts. For example, one study found that scaling exponents for population and number of incoming commuters are higher in developing countries, such as Brazil, than in the United States[30]. Thus, untangling the effects of population and connectivity should consider the myriad of physical and digital networks connecting cities in different geographic contexts. Additionally, current scaling models with connectivity inputs focus primarily on interpreting scaling exponents, such as how a city's position in an intercity network change different urban outputs' returns to scale[29,30,31]. Such approach assumes that intercity connectivity plays a similar role across cities and thus downplays the valuable heterogeneity in urban systems which can inform policy makers and planners about strategies and directions for improvements.

Here, we combine the urban scaling framework with data on social media, mobility, and scientific collaboration networks for hundreds of cities in China and the U.S. to explore whether intercity networks contribute to our understanding of urban performance. Our analysis focuses on analyzing the proportion of the scaling variance (heterogeneity in the scaling of patents, denoted as $R^2$) that can be explained by intercity connectivity, as opposed to traditional metrics such as population, GDP, and education. Our findings reveal that that connectivity can improve the explanatory power of population-based models for innovation (i.e., patents), increasing $R^2$ by 9% (in the U.S.) to 26% (in China). We also found that the role of connectivity increases with the complexity[34, 35] of these activities, which we approximate using the date of introduction of a patent's technologies[5]. Notably, intercity connectivity plays a more robust role in explaining variance in China, even after controlling for population, GDP, and education, but its influence is weaker in the United States. This discrepancy may be attributed to the different types of urban networks driving innovation in these contexts. In China, models incorporating social media and mobility networks can better predict the number of more complex patents, whereas in the United States, models with scientific collaboration networks are more predictive of complex patents. These findings help connect urban scaling and intercity network research by showing that intercity networks and population can combine to explain a city's innovative output. They also underscore policy implications, suggesting that developing countries may benefit from prioritizing investments in infrastructure or initiatives aimed at enhancing digital and mobility connectivity, whereas developed countries may find value in fostering scientific collaborations to bolster innovation.

**Methods**

**Independent Variables: Intercity connectivity**

We estimate the weighted degree-centrality of cities in the U.S. and China in three networks: social media connections, mobility, and co-publication using the formula below:

$$C_D(P_i) = \sum_{k=1}^{N} a(P_i, P_k)$$

where $C_D$ is the weighted degree centrality of a city $P_i$, which is calculated as the total count of other cities $P_k$ that $P_i$ is connected to, weighted by the total connections between $P_i$ and $P_k$.

The social media and mobility networks in China and the mobility network in the U.S. are calculated as directed networks and the rest as undirected networks. Results with eigenvector centrality are reported in Supplementary Figure S1 and Table S2. The spatial unit in China is a municipal city ($n$=338, excluding Hong Kong, Macau, and Taiwan), and the spatial unit in the U.S. is a Metropolitan Statistical Area (MSAs, $n$=380).

*Social media connections*. China's social media network data were collected from Weibo (Chinese Twitter) in 2015. We collected a random sample of 16.6 million Weibo users and aggregated connections based on the city where each account was registered. The resulting network consists of 338 cities (nodes) connected by weights given by the number of people from one city following people from another city. U.S. social media network data comes from Facebook's publicly available social connectedness index (SCI)[24]. We used the SCI version updated as of August 2020 on the county level.

To protect proprietary information, the SCI index only reports the scaled values of total Facebook connections between two counties normalized by multiplying Facebook users in two counties. To derive the total Facebook connections between two counties, we approximate the Facebook users in a county by multiplying the county population (from U.S. Census) by the percentage of Facebook users per county reported in the paper[36]. We then aggregate the county-to-county connections on the MSA level. The derived Facebook connections between MSAs are not accurate on the absolute values. Yet, the relative strength between the connections is still accurate and thus will not impact our scaling interpretations.

*Human mobility*. China's mobility network data were collected from Tencent Map in 2016. Tencent Map tracks mobility through location-sharing services on mobile devices. This is a daily dataset containing the top ten origins and destinations, the flow volumes, and modes of transportation (air, train, car), for each prefecture city in China. We averaged the amounts of flows among 338 cities in 2016 to derive daily flows for mobility networks. U.S. mobility network data comes from a dataset organized by GeoDS Lab[37]. The authors provided estimations of total population flow between counties based on the SafeGraph data, a company that collects mobile phone users' visit trajectories. We use the 2019 weekly flow data (12 months) aggregated to the MSA level.

*Intercity co-publication*. Both China and the U.S.'s co-publication network data were collected from the Web of Science in 2017. We extracted the zip codes from 365,134 papers' author affiliations and matched 281,243 with municipal cities (available for 293 cities). Similarly, we processed 592,880 papers in the U.S. and matched 397,370 with MSAs (available for 357 MSAs). To construct the co-publication network, we count one or multiple authors from one city and collaborate with an author from another as one co-publication count between the two cities. The weights in the network represent the number of papers co-authored by people between two cities.

**Socioeconomic Variables: Population, GDP, Employment, Education**

Population, GDP, and employment are classic predictor and response variables in traditional urban scaling research, representing the size of cities and the city outcomes. We collected China's municipal cities' urban area (registered) population, GDP, employment, and percentage of the population with a Bachelor's Degree or above (i.e., approximated by the percentage of the population employed in information, finance, and research industries) from the China City Statistical Yearbook 2016. Since many Chinese metropolitan cities have a large floating population that is not counted in the registered population, we also reported results with resident population (i.e., registered and floating population in urban area) (see Supplementary Note 1, Fig. S4, S5, S7, S12, and Table S3). The same socioeconomic variables for U.S. MSAs in 2019 were collected from the U.S. Bureau of Economic Analysis (see Supplementary Note 2).

**Dependent Variable: Patents**

We measure city outcomes in innovation through patents. For China, we collected total patent counts per city from China National Intellectual Property Administration (CNIPA) in 2017. We complement this data with patents by technological classes from PATSTAT, which extracts data from the European Patent Office (EPO). CNIPA data are more comprehensive, covering all patents filed for protection in China, while EPO data only covers Chinese patents filed for protection in Europe. Thus, we use CNIPA data to report the total number of patents in China and EPO data to analyze scaling outcomes by technology class.

For the U.S., we download 2019 patent data from PatentViews. Patents are geolocated with inventors' counties, and we aggregate the numbers on MSAs. Patents with multiple (e.g., $n$) inventors will have a weight (count) of $1/n$.

The complexity of patents is measured as the average number of years since the technology subclasses in a patent were introduced (i.e., more recent technological classes are considered to be more complex and/or sophisticated)[5] (see comparisons with another complexity metric in Supplementary Table S6). Raw patent data are grouped in CPC and USPC classification systems. We converted them to National Bureau of Economic Research (NBER) categories to leverage the existing complexity measures.

As alternative measures for city outcomes, we also report scaling results for scientific publication by fields and industry sectors (see Supplementary Figure S10, S11).

**Scaling models**

Urban scaling research focuses on how a city's properties scale as a function of city size[1]. This is done by modeling a city's output $Y$ as a power law of population $X$.

$$Y = AX^\beta,$$

Which, when log-transformed results in a linear model of the form:

$$\log(Y) = \log(A) + \beta \log(X)$$

Using lowercase letters for log-transformed variables, we obtain:

$$model_{pop}: y = A + \beta x$$

The coefficient indicates whether the relationship between city size and the outcome is linear (β=1), superlinear (concave) (β>1), or sublinear (convex) (β<1), *pop* represents total populations.

Here we expand this model to include measures of a city's connectivity *k* in physical and digital networks. That is, we use a model of the form instead:

$$model_{pop+networks}: y = A + \beta x + \gamma_1 k_1 + \gamma_2 k_2 + \gamma_3 k_3,$$

where $k_1$, $k_2$, $k_3$ are measures of a city's centrality in a network of social media, mobility connectivity, and academic collaborations, and $\gamma_1$, $\gamma_2$, and $\gamma_3$ are their respective scaling coefficients. Variations of this model are used to examine the contribution of networks as a whole and by each type.

We applied model_pop and model_pop+networks to three sets of city outputs — GDP, employment, and patents — and further breakdown patents by technological classes. Unlike previous scaling research, which focuses on scaling exponent, we focus on the scaling variance $R^2$ to compare the explanatory power between model_pop and model_pop+networks. Here, we refer to scaling variance as the enhanced model predictability of patent count ($R^2$) when incorporating inputs such as intercity connectivity. We attribute the difference in the $R^2$ between the models with and without connectivity as the contribution of the connectivity variables after controlling for population.

**Results**

**The shape of intercity networks**

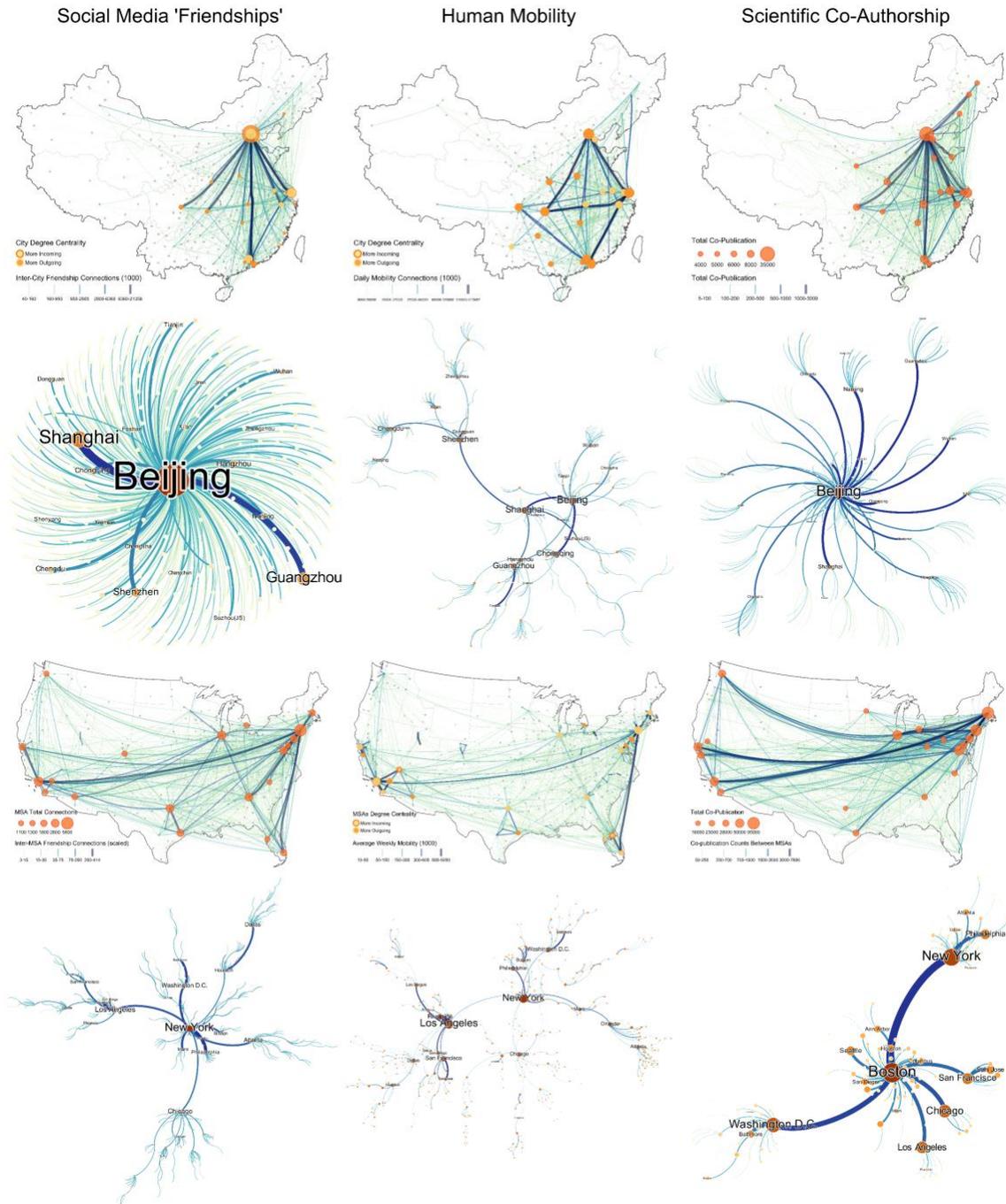

**Fig. 1 | Intercity networks in China and the United States**. The networks of scientific co-authorship, human mobility, and social media friendships for China and the United States, using both a geographic layout and the network's minimum spanning tree.

Figure 1 visualizes three key inter-city networks: social media, human mobility, and scientific collaborations for the United States and China. First, social media and co-publication networks seem to be more centralized than the physical mobility network and tend to be dominated by hub cities. This may reflect the fact that digital connections and researchers tend to concentrate in a few cities, taking advantage of long-distance connections that are no more costly than shorter ones. In China, the social media and co-publication networks center around Beijing, the country's cultural, political, and academic hub. In the United States, New York, and Boston play the role of Beijing in social media and scientific co-publication networks. Second, social media and co-publication networks in China and the U.S. involve connections among hubs (i.e., New York–Los Angeles, Boston–Chicago), whereas in comparison, mobility networks lack as many long-distance connections. As expected from a simple gravity model[38], mobility networks involve clusters centered on large cities, such as New York and Los Angeles in the U.S. or Shanghai, Shenzhen, and Beijing in China. These network properties support the idea that each of these networks plays a different role (see results of other metrics and top cities in each network in Supplementary Figure S1, S2).

**Models with connectivity inputs are better at predicting the number of patents**

Figure 2 shows how the total number of patents scales with a city's population and centrality using data from both China and the United States (see results of GDP and employment in Supplementary Figure S3 and results of resident population for China in Supplementary Figure S4). In all cases, we find a strong and positive correlation between population, connectivity, and city outcomes, as we should expect from the fact that population and connectivity are strongly correlated. Yet, we also find some meaningful differences.

While population is certainly an accurate predictor of total GDP and employment, its ability to fully explain a city's total number of patents is less clear. In China, population can only explain 56% of the variance in patents, considerably lower than the percentages observed between a city's population and its employment or GDP (respectively, $R^2=0.65$ and $R^2=0.71$). In the case of patents, the variance across cities correlates more strongly with a city's connectivity (social media $R^2=0.76$, mobility $R^2=0.70$ co-publication $R^2=0.64$) than population ($R^2=0.56$). Data from the U.S. show a similar pattern: population explains 96% and 98% of the variance in GDP and employment but only 69% in the total number of patents, which is lower than the 74% of variance explained using social media degree centrality. This finding hints at the fact that a city's position in the intercity network may account for variance in innovative activities that is unexplained by population, consistent with prior work[32].

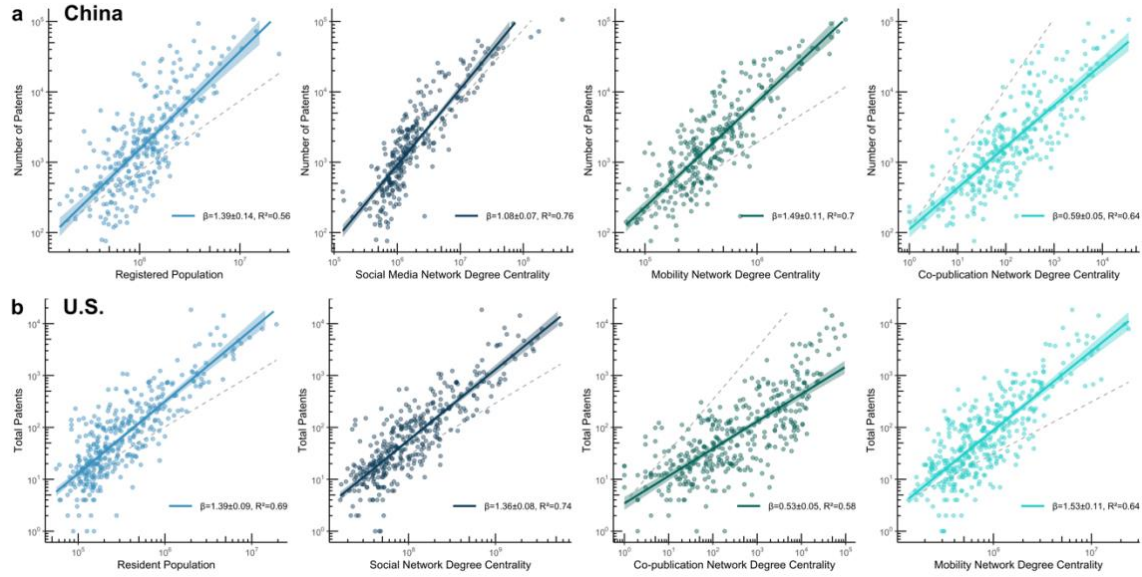

**Fig. 2 | The scaling results of population or degree centrality in the three networks against total patents**. **a**, China and **b**, the United States. Only cities that have values in the corresponding predictor and outcome variable are shown in each graphic.

**Table 1.** OLS regression results of each variable's coefficient estimate and its statistical significance.

| | OLS Regression: Socioeconomic Indicators and Connectivity versus Patents | | | | | | | | | |
|---|---|---|---|---|---|---|---|---|---|---|
| | China: Patents (log10) | | | | | U.S.: Patents (log10) | | | | |
| *Predictors* | M1: Est. | M2: Est | M3: Est. | M4: Est. | M5: Est. | M1: Est. | M2: Est | M3: Est. | M4: Est. | M5: Est. |
| Pop. (log10) | 1.38*** | 0.40** | | 0.15 | -0.16 | 1.38*** | 0.22 | | 0.04 | -0.21 |
| GDP (log10) | | 0.74*** | | | 0.33*** | | 0.84*** | | | 0.86*** |
| Pct w. Bachelor's (%) | | 0.07*** | | | -0.04* | | 0.05*** | | | 0.04*** |
| SM $C_D$ (log10) | | | 0.59*** | 0.55*** | 0.48*** | | | 0.89*** | 0.86*** | 0.17 |
| Mob $C_D$ (log10) | | | 0.46*** | 0.43*** | 0.50*** | | | 0.18 | 0.17 | 0.19 |
| Pub $C_D$ (log10) | | | 0.16*** | 0.14*** | 0.12** | | | 0.20*** | 0.20*** | 0.08*** |
| Observations | 275 | 275 | 275 | 275 | 275 | 366 | 366 | 366 | 366 | 366 |
| $R^2$ / $R^2$ adjusted | 0.55 | 0.72 | 0.80 | 0.81 | 0.82 | 0.69 | 0.83 | 0.78 | 0.78 | 0.84 |

*Note:* population alone (*M1*), population, GDP, and education (*M2*), networks alone (*M3*), population and networks (*M4*), and all-in-one (*M5*). Only cities that have values in all predictors are included. * *p<0.05*   ** *p<0.01*   *** *p<0.001*

We explore this idea further by comparing the result of predicting GDP and patents with five sets of inputs: population alone (*M1*), population with GDP and education (*M2*), networks alone (*M3*), population and networks (*M4*), and all-of-the-above (*M5*). The results are shown in Table 1. We find that while the model with population and network inputs (*M4*) contributes marginally over the model with population alone (*M1*) to explain GDP (from $R^2=0.71$ to $R^2=0.85$ for China; from $R^2=0.96$ to $R^2=0.97$ for the U.S.; see GDP results in Supplementary Table S1), it provides a better fit for the patent data ($R^2=0.81$ for China; $R^2=0.78$ for the U.S.; Both F-test $p<0.001$). When compared with models with population alone (*M1*), models with intercity connectivity inputs (*M4*) improve explanatory power ($R^2$) by 9% and 26% for the U.S. and China respectively. In China, even with the control of GDP and education, the complete model (*M5*) still increases $R^2$ by 10% as compared with the model with population, GDP, and education (*M2*) (F-test $p<0.001$).

In fact, in both U.S. and China, network models with and without population (*M3* and *M4*) report similar $R^2$, indicating that population and networks may behave as substitutes for each other. In addition, in China, all three network variables have statistically significant and positive effects on predicting the patents (*M4*), even after controlling for cities' GDP and education (*M5*), confirming that each network has a unique contribution. This observation is also robust with resident population (see Supplementary Table S3). In the U.S., only scientific collaboration connectivity contributes to explaining patterns after controlling for MSAs' GDP and education (*M5*), while the effects of social media and mobility networks are less clear. The model of population and networks (*M4*) actually performs worse than the model of population, GDP, and education (*M2*), indicating that networks are worse predictors than traditional socioeconomic indicators. Results using eigenvector centrality rather than degree centrality as the network measures are similar, except that the co-publication network loses its effect in China (see Supplementary Table S2). The difference between China and the U.S. again highlights the different roles that network connectivity may play in different economies.

One way that connectivity may help explain the scaling variance of patents is that it helps cities produce more patents than they're not-so-connected counterparts. Figure 3 compares the residuals of the population scaling model between high-connectivity and low-connectivity cities of a similar population (respectively colored in orange and purple) (see results with China's resident population in Supplementary Figure S5). Overall, we find that high-connectivity cities tend to be those that outperform in patenting activity in a population scaling model. The difference in China is more significant and consistent across population size but is still observed in the U.S. In China, the biggest mean and median difference in patents between high- and low-connectivity cities are observed at a population size of around 1-2 million (medium size), while for the U.S., big differences are observed in cities with a population of 10K-30K and 3-5 million (see T-tests for mean difference at Supplementary Table S4, S5). For example, Wenzhou, a city known for its entrepreneurial legacy (registered pop. 1.7m, resident pop. 2.3m) produces 45,385 patents in a year. In contrast, Guigang, a Chinese inland city in a low socioeconomic region with similar population (registered pop. 2m, resident pop. 2m), only produces 633 patents. Population cannot explain such differences without considering their differences in connectivity. Wenzhou has a high degree of centrality in social media and mobility networks: it is a hub for migrant workers in manufacturing industries. It has express trains connecting big metropolitans like Shanghai and Hangzhou as well as nearby manufacturing hubs such as Jinhua and Taizhou. Guigang, on the other hand, underperforms in all three networks. Similar contrasts can be found in U.S. cities, such as San Francisco-Oakland-Berkeley, CA, a well-known tech hub connected by numerous highways to the greater bay area and UC-Berkeley (pop. 4.7m, patent 14K), versus Riverside, which mainly has light-industry despite a strong mobility connection to the greater Los Angeles area (pop. 4.6m, patent 4K).

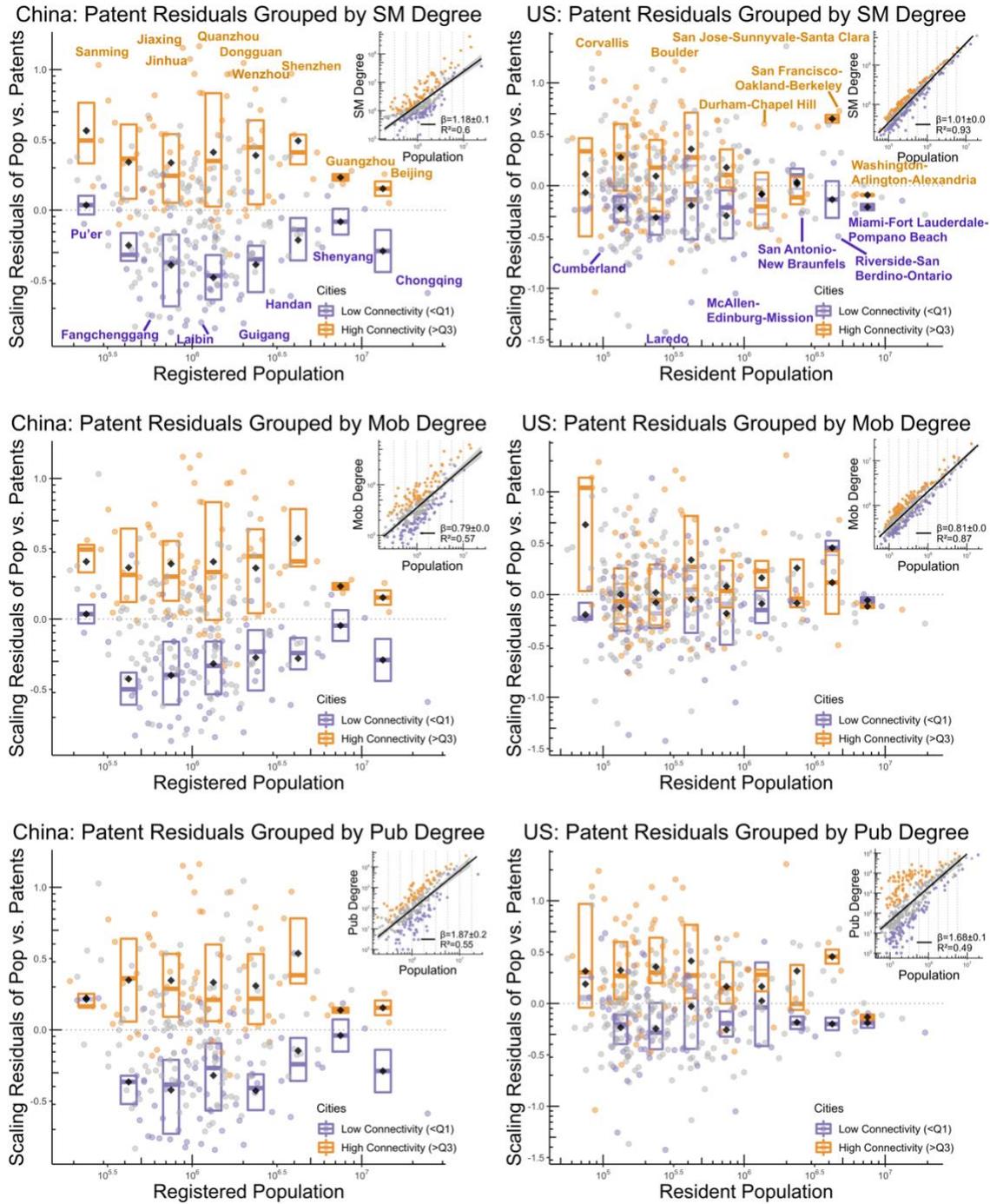

**Fig. 3 | Boxplots of scaling residuals for cities grouped by high/low connectivity and equal population intervals in a) China and b) U.S**. The Y axis is the residuals of patent scaling against population, which is a transformation of Figure 2. The boxplots show the range between the 1st and 3rd quantile, the median, and the mean (black) for high/low connectivity cities. The inset maps show high/low connectivity cities at each population interval. Orange dots are cities with high connectivity, defined as having a degree-

centrality above the 3rd quantile in a population interval. Purple dots are cities with low connectivity, defined below the 1st quantile (purple). Grey dots are cities with connectivity between 1st and 3rd quantile. The median splits population intervals with less than five dots. Small-to-medium size cities whose economy depends on tourism are removed. Cities labeled are examples with a similar population but the distinct outcome in patents.

After controlling for similar population, GDP, and education levels (see Figure 3 with GDP and education controlled in Supplementary Figure S6, S7 and matching analysis in Supplementary Figure S8), our findings stand (though to a lesser extent) and are robust for China (esp. medium-size cities), but not very significant for the United States. For example, in China, Jiaxing (pop. 0.9m, resident pop. 1.8m, patent 18K) is a tourist city in Zhejiang province and a manufacturing hub for printing equipment, textile, and chemical products, while Xining (pop. 1m, resident pop. 1.5m, patent 1K) in Qinghai province, whose main industries are renewable energy and services, is ten times lower in patent counts. These two cities also have an equal level of GDP (97b) and proportion of population working in information, finance, and research industries (3.6% and 3.8%). Such disparity may be attributed to Jiaxing's high connectivity: it has four times more social media followers and two times more mobility visits (yet only half of publication collaborations) than Xining. In the United States, intercity networks support some cities' innovation but not others. For example, Trenton-Princeton, NJ (pop. 0.4m, GDP 36b, college edu. 30%, patent 471), the capital city of New Jersey state and home to Princeton University, beats its less connected counterpart, Portland-South Portland, ME (pop. 0.5m, GDP 33b, college edu. 30%, patent 94) where many bank headquarters and an oil port are located, with four times more patents. Trenton-Princeton is also nearly ten times higher in scientific collaboration connections than Portland-South Portland, ME, and 8% and 48% higher in social media and mobility connections, respectively. In contrast, Baltimore-Columbia-Towson, MD has a similar population, GDP, and education profiles as Denver-Aurora-Lakewood, CO (pop. 3m, GDP 220b, and college edu. 30%) and is 20% higher in mobility connections and three times higher in scientific collaborations, but produces 62% (about 700) fewer patents than Denver.

We acknowledge that many other factors endogenous to high connectivity may also contribute to the disparity in cities' patent counts, such as the presence of prestigious research universities, the agglomeration of STEM industries and inventors, and accessible geography. We found that many overperformers are coastal cities, cities adjacent to metropolitans, and tech clusters (e.g., San Jose and Shenzhen) (see the geographic distribution of high/low connectivity cities in Supplementary Figure S9). In contrast, small-to-medium cities whose economy depends on tourism may also have exceptional connectivity, especially in mobility, but produce few patents. Though we cannot isolate these factors' effects from those of connectivity, they reveal the interactions between connectivity and local conditions.

**Models with connectivity inputs are better at predicting more complex patents**

Figure 3 shows that well-connected cities with STEM industries tend to have an edge on patent production and vice versa. We explore this point further by looking at scaling variance across patents of different vintages (patents making claims over technologies that appeared for the first time in a given year in the patent classification system). Figures 4a and 4c show that, in China and U.S., models with population and network inputs (i.e., $model_{pop+networks}$) show higher scaling $R^2$ than models with population alone (i.e., $model_{pop}$). As the complexity of patents increases, connectivity's explanatory power and its advantage over the population also increases, indicating that connectivity is a stronger predictor than population at predicting the number complex patents. For example, in China, the population is equivalent at predicting the number of patents in Communication versus Furniture and House Fixtures (11%), even though the former is more complex than the latter. With the addition of connectivity, the model can capture up to 64% of variances in Communication patents, which strongly signals that intercity connections contribute to or are affected by this type of innovation. Other fields that connectivity contributes the most to reduce variance include Computer Hardware & Software (47% increase in $R^2$), and Optics (44% increase in $R^2$). In the U.S., the addition of networks improves the scaling $R^2$ of patents in Measuring & Testing by 9%, Computer Hardware & Software by 8%, and Biotechnology by 7%. We also observed a similar pattern with industry and publication breakdown and with resident population in China (see Supplementary Figure S10, S11, and S12).

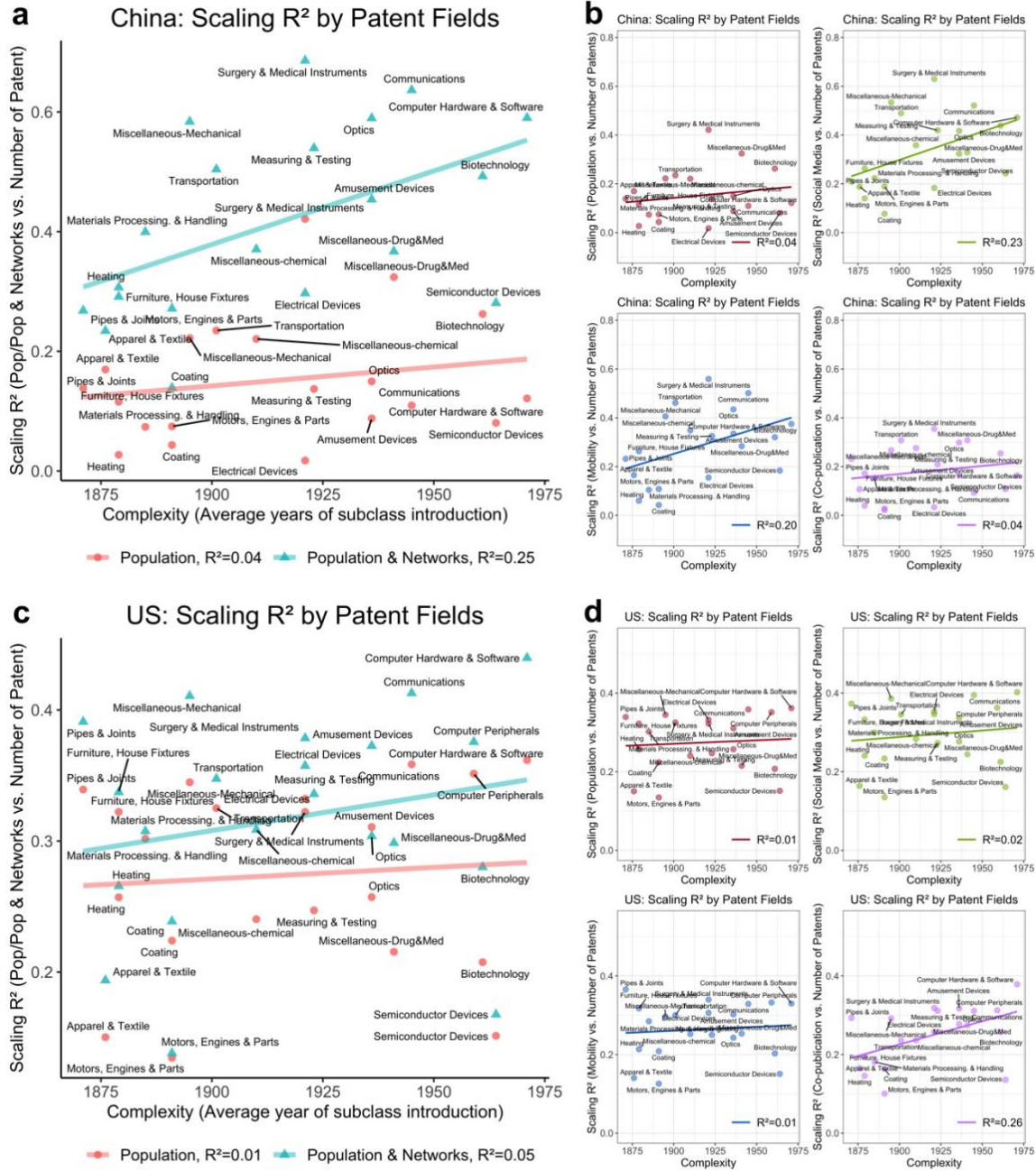

**Fig.4 | Comparing Scaling $R^2$ of model$_{pop}$ and model$_{pop+networks}$ by patent fields. a**, China **c**, U.S. The Scaling $R^2$ is further broken down using each independent variable in the model to examine each of its contribution in China (**b**) and U.S. (**d**). Small-to-medium size cities whose economy depends on tourism are removed.

Figures 4b and 4d further examine the findings by looking at how each network may reduce scaling variance. We found that the overall pattern persists using any network as input, but the ability to explain more complex patterns is more explicit with the social media and mobility networks in China while more explicit with the co-publication network in the United States.

**Discussion**

This paper attempted to untangle the role of population and intercity connectivity on city outcomes. We found that intercity connectivity sometimes explains the scaling variance of innovation (patents) better than population, but not the scaling variance of GDP and total employment. By observing outliers, we found that cities with high connectivity tend to outperform similar-sized cities with lower connectivity. This supports the idea that, in addition to population agglomeration effects, intercity connectivity may contribute to explaining variance in cities' innovation outcomes. We also found that intercity connectivity is a stronger predictor for more complex patents (i.e., patents of more recent vintage) whose production benefits the most from the flows of knowledge and information. Yet, these findings seem to hold stronger for medium-sized cities in China and are not significant after controlling for GDP and education in the U.S.

Should we expect cities in China and the United States to be impacted differently by connectivity? One possibility could be that networks in China and the U.S. play different roles. China has excellent communication (e.g., 5G technology) and transportation (e.g., high-speed rails) infrastructures which could enhance the role of networks in China compared to the United States. This infrastructure, combined with the fact that China has most of its population concentrated on the east, and within a single time zone (while the U.S. has two far-away coasts), could mean that China operates more as a single unit and the U.S. more as a collection of regional units (with the Northeast corridor, the Midwest, Texas, and the West Coast, acting more independently). In the U.S., patent innovation correlates strongly with intercity scientific collaborations, suggesting a key role for universities. These universities, coupled with a long-term sorting of industries and policy incentives, have enabled cities in tech clusters such as Silicon Valley (San Francisco and San Jose) or Research Triangle (Raleigh-Cary and Durham-Chapel Hill) to overperform on patent production.

Our findings suggest that intercity connectivity not just affects urban scaling by attracting educated and talented laborers to large metropolitans (selection effect)[29]; they also suggest connectivity contributes to urban innovation via social media influence, population mobility, and knowledge exchange (academic collaboration), even after we take into account the effects of city size, GDP, and education in China. These mechanisms correlate with population composition but have nuanced differences. For example, highly connected metropolitan cities in China tend to have a large floating population (or migrant workers) among their residents. This population is not highly educated and mostly works in manufacturing industries, but accounts for much of the population mobility and social media connections in China. We observed that some cities (e.g., Dongguan, a city known as the World's Factory in China) stand out less in innovation when we used resident population (including floating population) instead of registered population as its measure for city size, indicating that floating population of these cities may contribute to urban innovation. Yet, intercity connectivity still explains innovation differences in many other Chinese cities with matching resident population (and other socioeconomic variables), suggesting that intercity connectivity can accelerate innovation through other means beyond endogenous population composition.

Our study, however, has important limitations. First, we modeled population, intercity connectivity, and city outcomes in a simple linear scaling model, which could be expanded to include other variables and interaction terms. For example, intercity connectivity may help cities "borrow population" from their connected partners[39] and thus may better be considered as the population's exponent or coefficient rather than an independent term. Second, we noted that very few cities excel or fall behind in all three networks (20% in China and 7% in the U.S.). We found mixed evidence on whether improving a city's position in one network would improve its patent counts more than others (see Supplementary Figure S13). Future studies can investigate how a city's mismatched positions in different networks are associated with its innovation and industry compositions. Third, we observed some preliminary differences in the role of intercity connectivity in developing (i.e., China) vs.

developed (i.e., U.S.) countries. Still, more research is needed to examine how the patterns may be generalized to countries with different levels of development and local contexts and what causes the difference. Lastly, we provide robust correlational evidence but no causal evidence. In fact, intercity connectivity could be endogenous to innovation, since innovation could make a city more attractive increasing connections in social media, transportation, and mobility networks. Isolating the causal effects of connectivity on innovation is a difficult but important challenge for future research. Moreover, we did not separate the effects of industry compositions, geography, and local conditions, which are endogenous to connectivity. Qualitative approaches and case studies can complement our work to reveal how connectivity may impact innovation.

Our findings provide a first step connecting the urban scaling literature with intercity networks. This question is particularly of interest in a post-COVID world, as the pandemic has shifted the geography of some knowledge-intensive work (e.g., remote software developers). For example, the finding that small-to-medium size cities can better leverage their connectivity may be consistent with the post-COVID trend of people moving out of dense metropolitan areas to work-from-home locations[40]. This suggests that future cities that wish to become more competitive in innovation may borrow lessons from their more connected counterparts. Different networks also imply different kinds of policy interventions that local governments can enact to help the cities grow. For example, attracting high-impact industry leaders, developing physical infrastructure between cities such as road, train, and air networks, and elevating the prestige of local universities may correspond to better positions in social media, mobility, and scientific collaboration networks. Our research opens a new pathway for policymakers to rethink their strategies to actively improve their positions in one or more networks to uplift their cities' innovative outcomes[41].